 \documentclass[twocolumn,trackchanges]{aastex63} 
\usepackage[version=4]{mhchem}
\usepackage{ragged2e}
\usepackage{comment} 
\usepackage{float}
\usepackage{lipsum}

\newcommand{\cmq}{cm$^{-3}$~}

\newcommand{\kms}{\mbox{km s$^{-1}$}}

\newcommand\ee[1]{\ensuremath{\times 10^{#1}}}
\newcommand\e[1]{\ensuremath{ 10^{#1}}}
\newcommand{\msun}{M$_\odot$}
\newcommand{\rsq} {{$r^{-2}$~}}
\newcommand {\similar}{$\sim$~}
\newcommand{\approxt}{$\approx~$}
\received{XXXXX}
\revised{XXXXX}
\accepted{\today}
\submitjournal{ApJ}

\shorttitle{PSC in Orion}
\shortauthors{Sahu et al.}

\begin{document}


\title{ Density Structure of Centrally Concentrated Prestellar Cores from Multi-scale Observations}

 \correspondingauthor{Dipen Sahu}
 \email{dsahu@prl.res.in}

\author[0000-0002-4393-3463]{Dipen Sahu}
\affiliation{Physical Research laboratory, Navrangpura, Ahmedabad, Gujarat 380009, India}
\affiliation{Academia Sinica Institute of Astronomy and Astrophysics, 11F of AS/NTU Astronomy-Mathematics Building, No.1, Sec. 4, Roosevelt Rd, Taipei 10617, Taiwan, R.O.C.}

\author[0000-0012-3245-1234]{Sheng-Yuan Liu}
\affiliation{Academia Sinica Institute of Astronomy and Astrophysics, 11F of AS/NTU Astronomy-Mathematics Building, No.1, Sec. 4, Roosevelt Rd, Taipei 10617, Taiwan, R.O.C.}

\author[0000-0002-6773-459X]{Doug Johnstone}
\affiliation{NRC Herzberg Astronomy and Astrophysics, 5071 West Saanich Rd, Victoria, BC, V9E 2E7, Canada}
\affiliation{Department of Physics and Astronomy, University of Victoria, Victoria, BC, V8P 5C2, Canada}

\author[0000-0002-5286-2564]{Tie Liu}
\affiliation{Shanghai Astronomical Observatory, Chinese Academy of Sciences, 80 Nandan Road, Shanghai 200030, People's Republic of China}

\author[0000-0001-5175-1777]{Neal J. Evans II}
\affiliation{Department of Astronomy The University of Texas at Austin 2515 Speedway, Stop C1400 Austin, TX 78712-1205, USA}

\author[0000-0001-9304-7884]{Naomi Hirano}
\affiliation{Academia Sinica Institute of Astronomy and Astrophysics, 11F of AS/NTU Astronomy-Mathematics Building, No.1, Sec. 4, Roosevelt Rd, Taipei 10617, Taiwan, R.O.C.}

\author[0000-0002-8149-8546]{Ken'ichi Tatematsu}
\affiliation{Nobeyama Radio Observatory, National Astronomical Observatory of Japan,
National Institutes of Natural Sciences,
462-2 Nobeyama, Minamimaki, Minamisaku, Nagano 384-1305, Japan}

\author[0000-0002-9289-2450]{James Di Francesco}
 \affiliation{NRC Herzberg Astronomy and Astrophysics, 5071 West Saanich Rd, Victoria, BC, V9E 2E7, Canada}
\affiliation{Department of Physics and Astronomy, University of Victoria, Victoria, BC, V8P 5C2, Canada}
 
\author[0000-0002-3024-5864]{Chin-Fei Lee}
\affiliation{Academia Sinica Institute of Astronomy and Astrophysics, 11F of AS/NTU Astronomy-Mathematics Building, No.1, Sec. 4, Roosevelt Rd, Taipei 10617, Taiwan, R.O.C.}

\author[0000-0003-2412-7092]{Kee-Tae Kim}
\affiliation{Korea Astronomy and Space Science Institute, 776 Daedeokdae-ro, Yuseong-gu, Daejeon 34055, Republic of Korea }
\affiliation{University of Science and Technology, Korea (UST), 217 Gajeong-ro, Yuseong-gu, Daejeon 34113, Republic of Korea}

\author[0000-0002-2338-4583]{Somnath Dutta}
\affiliation{Academia Sinica Institute of Astronomy and Astrophysics, 11F of AS/NTU Astronomy-Mathematics Building, No.1, Sec. 4, Roosevelt Rd, Taipei 10617, Taiwan, R.O.C.}

\author[0000-0002-1369-1563]{Shih-Ying Hsu}
\affiliation{National Taiwan University (NTU), Taiwan, R.O.C.}
\affiliation{Academia Sinica Institute of Astronomy and Astrophysics, 11F of AS/NTU Astronomy-Mathematics Building, No.1, Sec. 4, Roosevelt Rd, Taipei 10617, Taiwan, R.O.C.}

\author[0000-0003-1275-5251]{Shanghuo Li}
\affiliation{Max Planck Institute for Astronomy, K\"onigstuhl 17, 69117 Heidelberg, Germany}

\author[0000-0003-4506-3171]{Qiu-Yi Luo}
\affiliation{Shanghai Astronomical Observatory, Chinese Academy of Sciences, 80 Nandan Road, Shanghai 200030, People's Republic of China}

 \author[0000-0002-7125-7685]{Patricio Sanhueza}
\affiliation{National Astronomical Observatory of Japan, National Institutes of Natural Sciences, 2-21-1 Osawa, Mitaka, Tokyo 181-8588, Japan}
\affiliation{Department of Astronomical Science, SOKENDAI (The Graduate University for Advanced Studies),
2-21-1 Osawa, Mitaka, Tokyo 181-8588, Japan}

\author[0000-0001-8385-9838]{Hsien Shang}
\affiliation{Academia Sinica Institute of Astronomy and Astrophysics, 11F of AS/NTU Astronomy-Mathematics Building, No.1, Sec. 4, Roosevelt Rd, Taipei 10617, Taiwan, R.O.C.}

\author[0000-0003-1665-6402]{Alessio Traficante}
\affil{IAPS-INAF, via Fosso del Cavaliere 100, I-00133, Rome, Italy}

\author[0000-0002-5809-4834]{Mika Juvela}
\affiliation{Department of Physics, P.O.Box 64, FI-00014, University of Helsinki, Finland}

\author[0000-0002-3179-6334]{Chang Won Lee}
\affiliation{Korea Astronomy and Space Science Institute, 776 Daedeokdae-ro, Yuseong-gu, Daejeon 34055, Republic of Korea }
\affiliation{University of Science and Technology, Korea (UST), 217 Gajeong-ro, Yuseong-gu, Daejeon 34113, Republic of Korea}

\author[0000-0002-5881-3229]{David J. Eden}
\affiliation{Astrophysics Research Institute, Liverpool John Moores University, iC2, Liverpool Science Park, 146 Brownlow Hill, Liverpool, L3 5RF, UK.}

\author[0000-0002-6622-8396]{Paul F. Goldsmith}
\affiliation{Jet Propulsion Laboratory, California Institute of Technology, 4800 Oak Grove Drive, Pasadena, CA 91109, USA}

\author[0000-0002-9574-8454]{Leonardo Bronfman}
\affiliation{Departamento de Astronomía, Universidad de Chile, Camino el Observatorio 1515, Las Condes, Santiago, Chile}

 \author[0000-0003-4022-4132]{Woojin Kwon}

\affiliation{Department of Earth Science Education, Seoul National University, 1 Gwanak-ro, Gwanak-gu, Seoul 08826, Republic of Korea}
\affiliation {SNU Astronomy Research Center, Seoul National University, 1 Gwanak-ro, Gwanak-gu, Seoul 08826, Republic of Korea}

\author[0000-0003-3119-2087]{Jeong-Eun Lee}
\affiliation{Department of Physics and Astronomy, Seoul National University, 1 Gwanak-ro, Gwanak-gu, Seoul 08826, Korea}

 \author {Yi-Jehng Kuan}
\affiliation{Department of Earth Sciences, National Taiwan Normal University, Taipei, Taiwan (R.O.C.)}
\affiliation{Academia Sinica Institute of Astronomy and Astrophysics, 11F of AS/NTU Astronomy-Mathematics Building, No.1, Sec. 4, Roosevelt Rd, Taipei 10617, Taiwan, R.O.C.}

\author{Isabelle Ristorcelli}
 \affiliation{ Universit\'e de Toulouse, UPS-OMP, IRAP, F-31028 Toulouse cedex 4, France}

\begin{abstract}
 Starless cores represent the initial stage of evolution toward (proto)star formation, and a subset of them, known as  prestellar cores, with high density (\similar \e6 \cmq or higher) and being centrally concentrated are expected to be embryos of (proto)stars. 
Determining the density profile of prestellar cores, therefore provides an important opportunity to gauge the initial conditions of star formation.  
In this work, we perform rigorous modeling to estimate the density profiles of three nearly spherical prestellar cores among a sample of five highly dense cores detected by our recent observations.
We employed multi-scale observational data of the (sub)millimeter dust continuum emission including those obtained by SCUBA-2 on the JCMT with a resolution of \similar 5600 au and by multiple ALMA observations with a resolution as high as \similar 480 au.
We are able to consistently reproduce the observed multi-scale dust continuum images of the cores with a simple prescribed density profile, which bears an inner region of flat density and a \rsq profile toward the outer region.
By utilizing the peak density and the size of the inner flat region as a proxy for the dynamical stage of the cores, we find that the three modeled cores are most likely unstable and prone to collapse.
The sizes of the inner flat regions, as compact as \similar 500 au, signify them  being the highly evolved prestellar cores rarely found to date.

\end{abstract}




\section{Introduction} \label{sec:intro}

Starless cores are  dense fragments in molecular clouds but without the association of (proto)stars.
A subset of starless cores that are gravitationally unstable, dubbed prestellar cores, will be prone to collapse and eventually form (proto)stars  \citep{Crapsi2005ApJ...619..379C, diFrancesco2007prpl.conf...17D}.
These prestellar cores (PSCs), therefore at the first stage of star formation, offer the best opportunity to gauge the initial conditions of star formation.
Characterizing their structures, such as density and temperature profiles, is essential for testing theoretical models and for understanding the physical processes of star formations \citep{Andre2000}.

Observationally, the density profile of starless/prestellar cores can be estimated from the dust continuum emission, the dust extinction, and/or emission from molecular lines. 
The latter method can be less accurate, as tracer species get depleted onto the dust grains at the low temperatures and high densities within the PSCs.
Using NIR extinction observations, \citet{Alves2001Natur.409..159A} estimated the density profile of the dark core B68. They found that the density profile closely fit with the theoretical description of the Bonner-Ebert (BE) sphere \citep{Bonnor1956, ebert1955ZA.....37..217E}, which represent a class of externally pressure bound self-gravitating isothermal cores.
In particular, B68 is found to be a `critical' BE sphere configuration, suggesting that the core is on the verge of instability.

Studies by \citet{Ward-Thompson1994, 1996A&A...314..625A, 1999MNRAS.305..143W} and \citet{2000A&A...361..555B} used millimeter dust continuum observations to estimate PSC density profiles, finding that the density profile of starless/prestellar cores can be fitted with power laws that are steeper in the outer region and shallower in the inner region. 
Based on similar observations, \citet{2002ApJ...569..815T} suggested that the density profile of starless cores can be better fit by the following equation: 
\begin{equation} \label{eq1} 
\rho(r)=\frac{\rho_c}{1 + (r/a)^\alpha} ,
\end{equation}
where $r$ is the radial distance and  $a$ is the radial size of the inner flat region and  $\alpha$ is the power law index. Now, with $\alpha$ =2 in Eq~\ref{eq1}, the underlying profiles resembles the BE sphere model. 
 \citet{Daap2009MNRAS.395.1092D} used this form of equation with a fixed power-law, $\alpha$ =2, and  emphasized that the generic features of  the BE 
sphere is not unique to equilibrium conditions but also appear in  gravitationally  collapsing objects too. Additionally the authors argued that the flat region size can be used as a proxy for core evolution. Estimation of the flat region size of prestellar cores is therefore important \citep[e.g., see also ][]{Keto2008ApJ...683..238K, keto2010MNRAS.402.1625K}.

When using optically-thin dust emission for characterizing the density structures of the prestellar cores and for inferring their evolution, multi-scale observations are indispensable.
Single-dish telescopes can image the full core scale but cannot provide sufficient angular resolution to capture the central zone.
Interferometric observations, typically resolve out the large scale structures of the starless core, but constrain the size of the inner flat region.
Indeed, for L1544, which is one of the most studied cores having observations icluding both single dish and interferometric \citep[e.g.,][]{Ohashi1999ApJ...518L..41O, Caselli2019ApJ...874...89C}, the derived column density profile based on single-dish observations varies as $r^{-1}$, with a flatter region within $r$ \similar 3000 au \citep{2002ApJ...565..331C}. Detailed fitting using an unstable BE sphere model required that the flat extent of L1544 is $\gtrsim$ 1500 au \citep{keto2010MNRAS.402.1625K,keto2014MNRAS.440.2616K}.

To date, only a few prestellar cores have been studied in detail \citep[e.g.,][]{Crapsi2004, Schmalzl2014} due to the limited number of candidate cores.
The lack of known PSCs is most likley due to a combination of source structure and limited sensitivity \citep[see introduction section of ][for more details]{sahu2021}. In particular, the structure of the innermost region within the PSCs  had been least studied \citep{Caselli2019ApJ...874...89C}.
Recently, \citet{sahu2021} detected five prestellar cores in Orion, which are centrally dense and have peak densities that are  quite high (10$^6$ - 10$^7$cm$^{-3}$)  relative to other known prestellar cores, such as L1544 ( $\sim 10^6$ \cmq),   and `starless' cores L1498 and L1517B \citep{Tafalla2004A&A...416..191T}.

The primary motive of the present work is to estimate the density profile from small ($\lesssim $1000 au) to large ($\sim10000$s au) scales toward the three comparatively spherical prestellar cores among the above five, namely G209.94-19.52N, G212.10-19 15N1, and G205.46-14.46M3 (hereafter G209N, G212N1, and G205M3, respectively).
We estimate the peak density and the size of flat regions of these PSCs consistently through multi-scale observations, and use these measurements as proxies for the cores' dynamical state. 

The paper is organized as follows; in Section 2 we describe the observations. In Sections 3 we introduce the physical model and the methodology. Results of the modeling are described in Section 4. The limitations and implications of the model results are discussed in Section 5.  Finally, in Section 6 we summarize our findings. Other relevant information is provided in the Appendix.

\section{Observations}\label{obs}

\subsection{Single dish observations}
As a part of the James Clerk Maxwell Telescope (JCMT) SCOPE survey (proposal code: M16AL003), 58 Planck Galactic Cold Clumps (PGCCs) were observed across the Orion\,A, Orion\,B, and $\lambda$-Orionis molecular clouds. The Submillimetre Common-User Bolometer Array 2 (SCUBA-2) instrument \citep{Holland2013MNRAS.430.2513H} at the JCMT was used to map the 850 $\mu$m dust continuum emission at an angular resolution $\sim$14\arcsec— details of the full dataset are described by \citet{2018ApJS..236...51Y}.
Figure~\ref{aca_scuba} uses gray contours to plot the JCMT 850 $\mu$m dust continuum emission  toward the three Orion cores G209N, G212N1, and G205M3.

\subsection{Interferometric observations: ALMA}

ALMASOP (project ID: 2018.1.00302.S) ALMA observations were carried out in Cycle 6 toward 72 fields selected from the above JCMT SCOPE survey. 
The observations were executed in the 1.3 mm band using three different array configurations: 12 m C43-5 (TM1), 12 m C43-2 (TM2), and 7 m ACA, resulting in angular resolutions ranging from 0\farcs34 to 5\farcs5.
The correlator was configured into four spectral windows with 1.875 GHz bandwidth each. 
We adopted a coarse velocity resolution of 1.129 MHz, equivalent to 1.5 \kms, to facilitate efficient continuum observations and maximize spectral line coverage. 
The remaining details of the observational parameters are presented in \citet{2020ApJS..251...20D}. 

\begin{figure*}
\centering
    \includegraphics*[width=17.5cm]{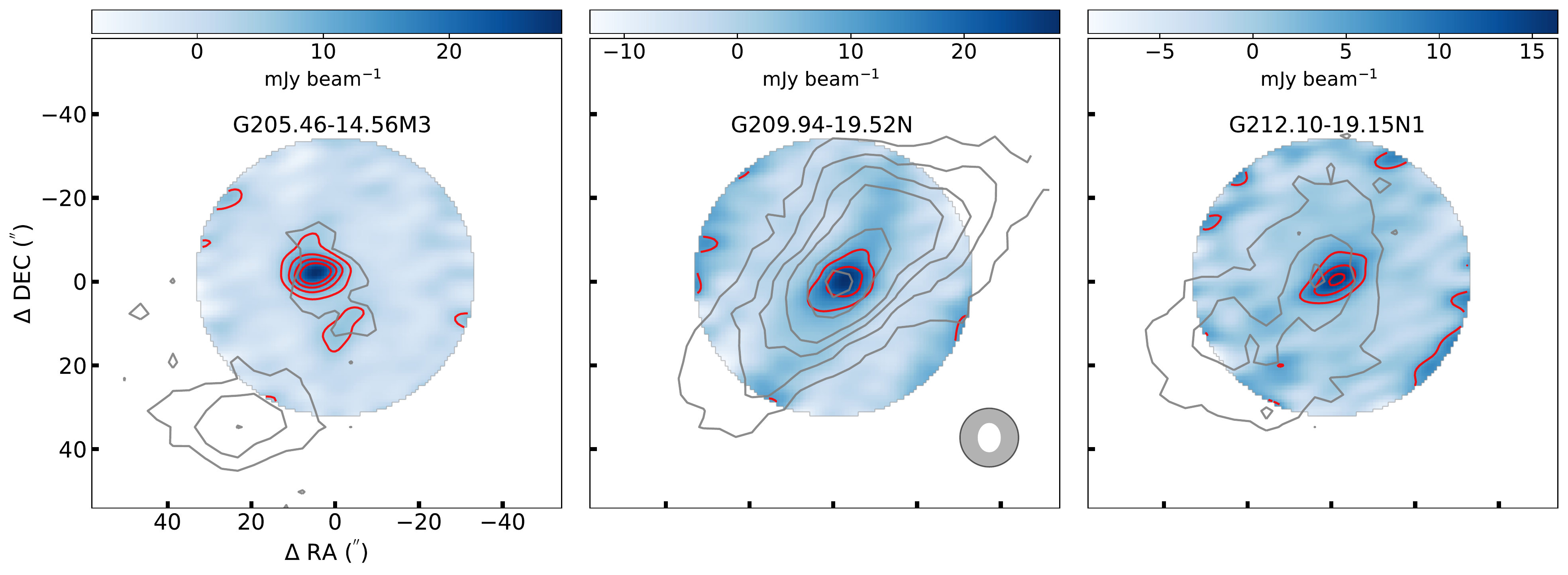}
    \caption{The 1.3 mm dust continuum emission of the three prestellar cores ( in blue colors) as observed with ACA-ALMA; red contours correspond to 5, 10, 15$\sigma$, where $\sigma$(rms) are 1.0,  2.0, 1.0 mJy~beam$^{-1}$ respectively. The gray contours correspond to 0.85mm SCUBA-2 emission with similar contour levels where  $\sigma$(rms) =44.5, 13.9, 16.4 mJy~beam$^{-1}$ respectively. The beams of SCUBA-2 and ACA are shown in the middle panel. 
} 
    \label{aca_scuba}
\end{figure*}

\begin{figure*}
    \centering
    \includegraphics*[width=17.5cm]{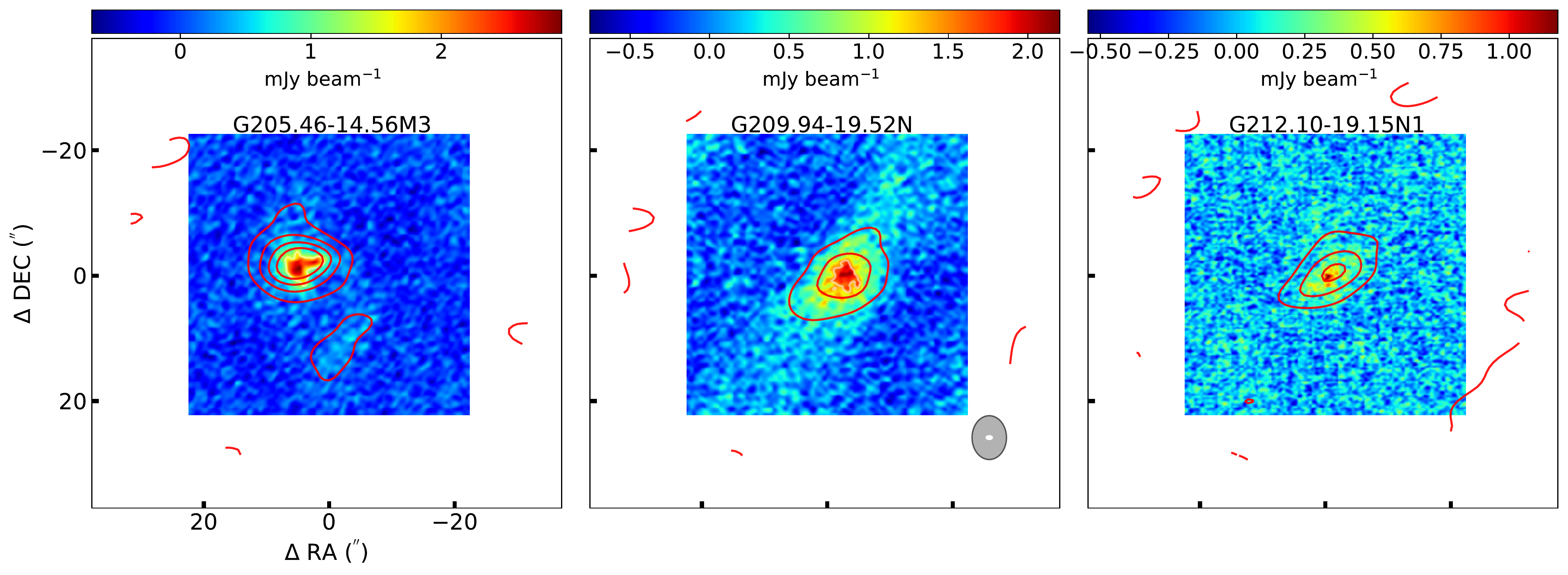}
    \caption{The 1.3 mm dust continuum emission of the three prestellar cores  - color scale from ACA+TM2 combined data ( resolution$\sim$1\farcs2 ). The red contours correspond to ACA observation similar to those shown in Fig.1. The beam sizes of ACA and TM2+ACA configuration  are represented in the middle panel. 
}
    \label{aca_tm2}
\end{figure*}

The visibility data were calibrated with the CASA 5.4 \citep[Common Astronomy Software Applications package;][]{McMullin2007} pipeline script as delivered by the observatory.
The visibility data for all configurations and executions toward each of the 72 fields were then separated into continuum and spectral lines and imaged jointly.
The 1.3 mm continuum images of each field were generated through CASA's {\it tclean} task with the ``automask" on, the hogbom deconvolver, and a robust weighting of 0.5 for data acquired with different arrays. 
The ACA-only 1.3mm continuum images toward the three cores, G209N, G212N1, and G205M3 are shown in Figure~\ref{aca_scuba} in blue, with red contours at a 5\farcs5 resolution. 
In addition, all three cores were also detected by the 12m array in configuration TM2 in combination with ACA at 1\farcs2 resolution.
Figure~\ref{aca_tm2} shows a magnified view of the cores using the high-resolution TM2 observations in color and ACA observations overplotted in contours. 
As noted in \cite{sahu2021}, these PSCs are resolved out by the 12m-TM1 array at a resolution \similar 0\farcs3.

\section{Physical model and methodology } \label{physical model}

For inferring the density profile of the observed prestellar cores, we adopt the parametric function proposed in \citet{Daap2009MNRAS.395.1092D} 

 \begin{equation} \label{eq2}
    \rho(r)  =
  \begin{cases}
    \frac{\rho_c ~a^2}{r^2 + a^2} & \text{for } r\leq R, \\
     0         & \text{for}  ~ r > R.
   \end{cases} 
   \end{equation}
Here $\rho_c$ is the  peak  volume density ($n_c$, corresponding number density), $a$ is the size of the inner flat region, and {\it R} is the outer radius of the core.

An analytical column density profile can be integrated from Eq.~\ref{eq2} as shown by \citet{Daap2009MNRAS.395.1092D}:

 \begin{equation}\label{eq3}
 \begin{aligned}
    \Sigma(x) &= \frac{\Sigma_c}{\sqrt{1+(x/a)^2}}  \times \\
 &  [{\rm arctan}(\sqrt\frac{c^2-(x/a)^2} {1 + (x/a)^2}) /{\rm arctan}(c) ]  + d,
\end{aligned}
\end{equation}
where $\Sigma_c$ is the peak column density, and c = R/a. The flat radius is given by, $a = k \frac{C_s}{\sqrt(G \rho_c)}$, where $C_s$ is the sound speed, $G$ is the gravitational constant, and $k$ is a free variable proportionality constant. 
The peak column density is related to the peak density via the relation: $\Sigma_c =2 a\,  n_c\, {\rm arctan}(c)$. 
In the equation above, we include an additional term $d$ to represent the gas column density due to the extended ambient cloud surrounding the prestellar core. 
It is to be noted that the underlying assumption of the Eq.~\ref{eq2} is isothermal condition. Subsequently, we assume a constant temperature and opacity through the cores. 
The above assumption works well with the scope of this study and we discuss further in Section~5 the possible implications.

\begin{figure*}[t!]
    \centering
    \includegraphics [width=16cm]{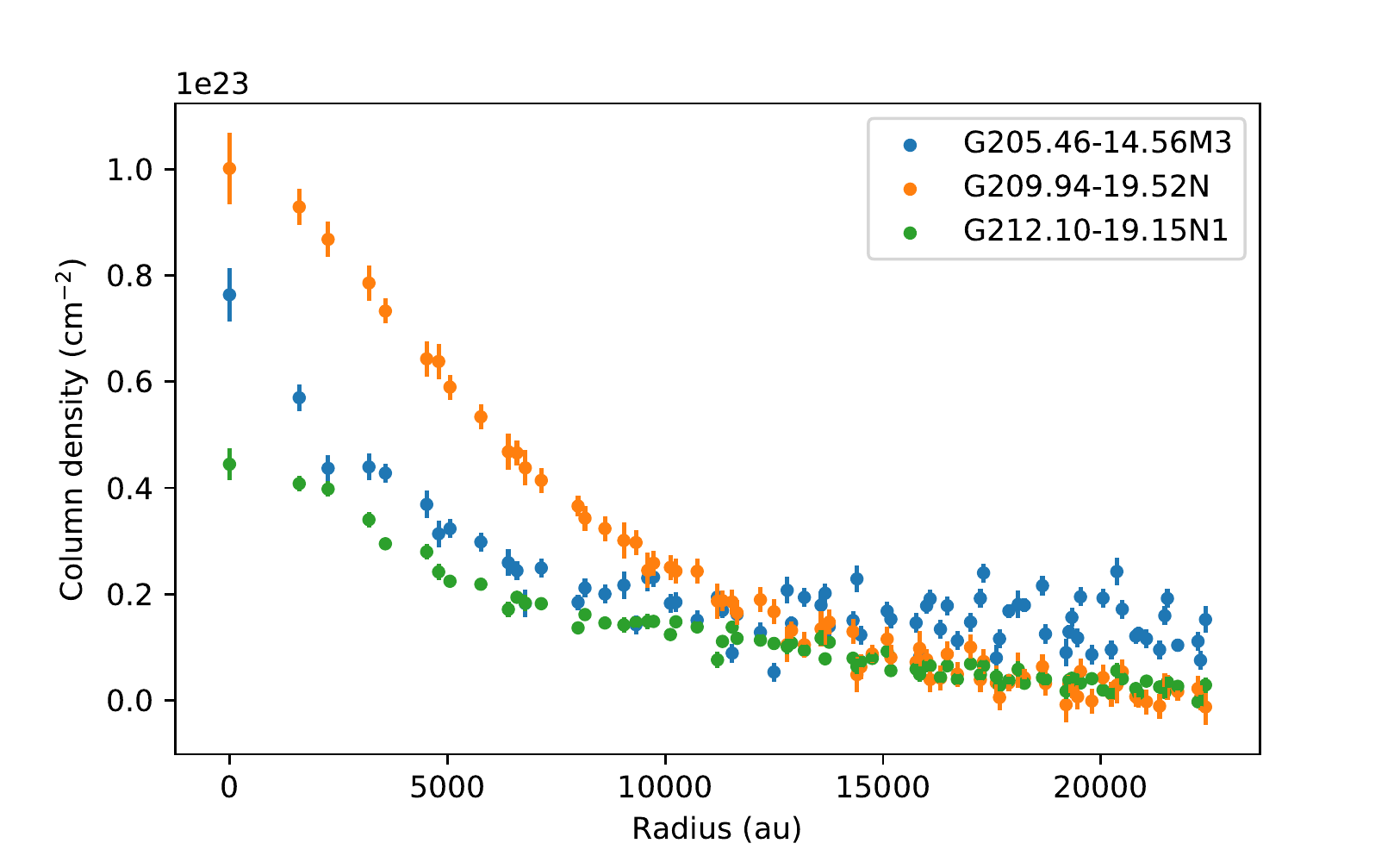}
    \caption{Azimuthally averaged column density profile of the three prestellar cores as obtained from SCUBA-2 dust continuum emission. The error bars in the profile are the statistical errors obtained by $\sigma/\sqrt(n)$, where $\sigma$ is rms noise obtained from the emission map, and `n' is the number of the pixels corresponding to the radial distance bin used to obtain the radially average flux values.}
   
    \label{radialprofiles}
\end{figure*}

To extract the underlying density profile of each observed core, we first estimate the column density profile from the annular-averaged dust continuum intensity profile at 850 {$\micron$} observed with the JCMT by assuming the emission is optically thin with a dust absorption coefficient ($\kappa$)\similar 2.23 $\mathrm{cm^{2}g^{-1}}$ and that the dust temperature is 10~K throughout (Fig.~\ref{radialprofiles}).
These profiles are then compared against the analytic column density profile (Eq.~\ref{eq3}) after the latter gets convolved with a Gaussian profile representing the JCMT beam FWHM $ \sim 14''$.
By fitting the radial profiles, we determine the outer boundary $R$ of each prestellar core and the constant offset $d$ due to the cloud.
We also determine a rough estimate of the peak density $\rho_c$ and the flat radius $a$ guiding the multi-scale imaging synthesis and fitting.

With the outer edge of each prestellar core $R$ determined, we generate a set of 2000 synthetic model 1.3~mm continuum images by varying $a$ with 40 linearly spaced values between 10 au and 4000 au and  $n_c$ with 50 logarithmic-spaced points between 10$^5$ \cmq and 10$^9$ \cmq, as input models for the radiative transfer code SPARX\footnote{https://sparx.tiara.sinica.edu.tw/}. A dust temperature of 10~K throughout the cores and a dust absorption coefficient of \similar 0.9 $\mathrm{cm^{2}g^{-1}}$ from \citet{1994A&A...291..943O} corresponding to thick icy dust at 10$^6$ \cmq are adopted.
The radiative transfer code, with fixed temperature, ensures that any non-negligible optical depth near the core centre is properly taken into account during the calculation.

The telescope-related interferometric effects  are further considered through imaging simulations using CASA.
We incorporate the same telescope configuration as in the original ALMA observations, process the above synthetic model images using the `simobserve' task, and produce visibility data sets for each of the 2000 synthetic models.
We use the `tclean/simanalyse' task, as we did for the real observed images, to further produce mock continuum emission maps, which can then be directly compared with the observed images.

We estimate the best fit parameters by comparing the radial intensity profiles between the observed and the simulated images using $\chi^2$ fitting. 
For each core, separate $\chi^2$ fittings were conducted for the images observed at multiple scales with the JCMT and the ALMA ACA, and TM2 configurations. 
The multi-scale approach is necessary as the core emission at large physical, and hence angular scales, is best detected by single-dish telescopes and not the interferometric observations. Contrariwise, the inner radial features, smaller than 2800~au in size, are poorly constrained by the JCMT SCUBA-2 observations, given the 14\arcsec ~beam size of SCUBA-2 at the distance of the Orion cores (\similar 400 pc).

We present the  $\Delta \chi^2$ plots corresponding to one sigma value ($\sigma$)  (see Appendix, Fig.~\ref{G205M3:dchi2}, \ref{G209N dchi2}, \ref{G212 dchi2}) for each set of observations in order to estimate consistent model parameters for the observed images at all scales.
The best parameters for minimizing $\chi^2$ over the individual maps and for the global optimization are given in Table 1.
We find that the flat radius $a$ of the cores is small, 300 - 1400~au, and describe below the results for each individual core in detail.

\section{Results}

The major aim of our multi-scale imaging modeling is to estimate the overall size of the prestellar core, $R$, the size of the inner flat region at the center of the core, $a$, and the associated peak density, $\rho_c$, using the assumed physical model (Eq.~\ref{eq2}). 
As the cores are found with $R/a > 10$, their masses are also approximated as $\mathrm{M \sim ~4 \pi ~\rho_c a^2 R}$, within an uncertainty of $<15\%$ \citep{Daap2009MNRAS.395.1092D}.
The size of the inner flat region acts as a proxy of prestellar core evolution and based on our estimation, the PSCs presented here possibly are the most evolved and compact (i.e., have smallest flat region) compared to the PSC samples found in the literature.
Below, we discuss the three PSCs individually.

\subsection{G205M3}
\begin{figure*}
    \centering
    \includegraphics[width=18cm]{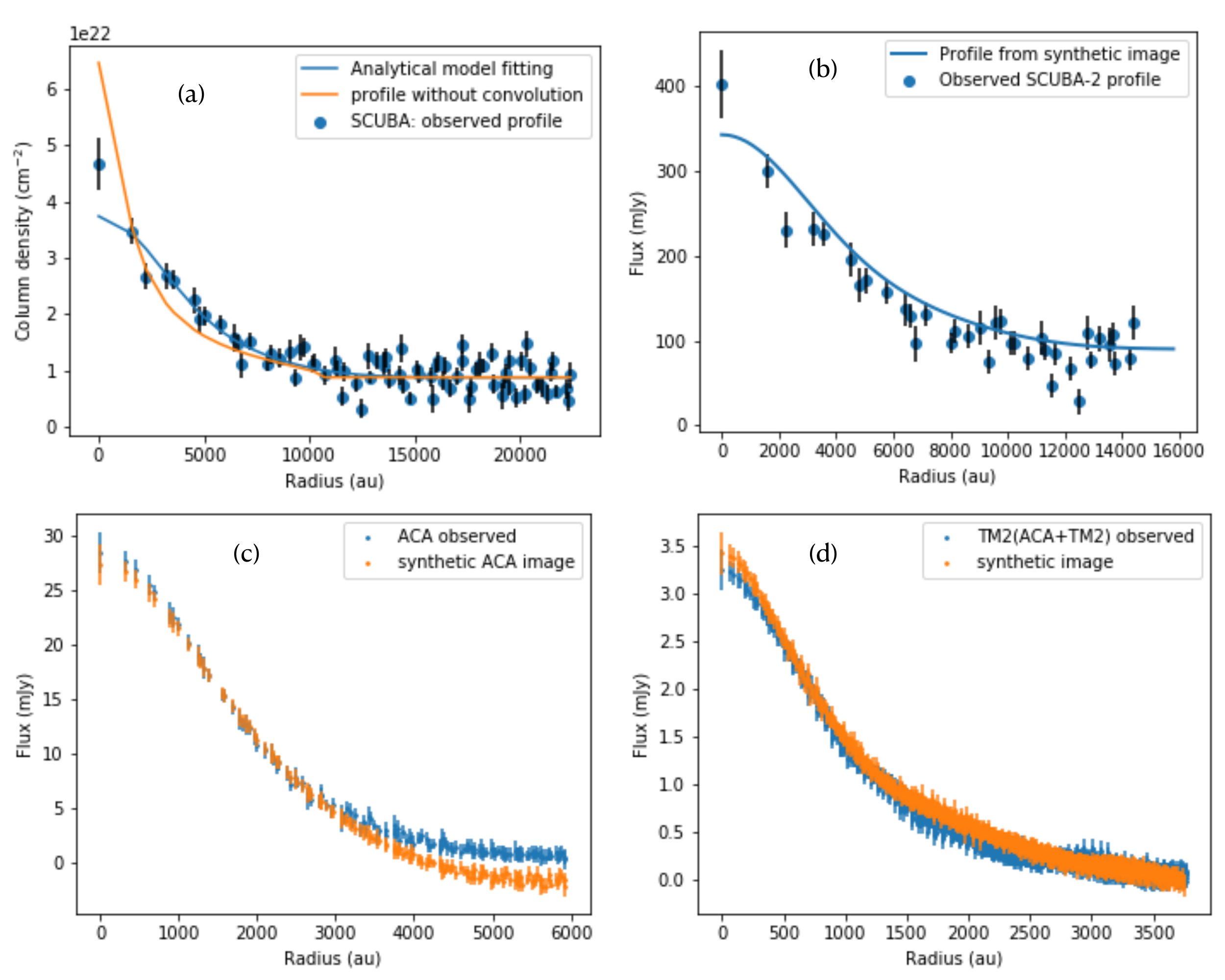}
    \caption{
    Radial column density profile fitting for G205M3 core. (a) Simple analytical model fitting to determine the extent of core radius. (b) Profile fitting for SCUBA-2 observation with the profile from the model images corresponding to the end row values of Table~\ref{table1}. (c) Similar to the case of b, but for ACA observations (d) similar to b,c for ACA+TM2 observations.  The model parameters are noted in Table~\ref{table1}. A comparison between the modeled images with the observed maps are shown in Appendix  Fig.~\ref{G205_modelmap}.)}
    \label{G205_fitting}
\end{figure*}

Among the sample prestellar cores, G205M3 is the only core that shows substructure at the 1000 au scale \citep[see Fig. 2 and ][for details]{sahu2021}.
The SCUBA-2 image (see Fig.~1) shows that the core is clearly separated from another nearby core to the SE.

Within the core, however, a fainter second peak, roughly 20\arcsec ~SW from center, can be seen in both the SCUBA-2 and ACA maps.
Despite the substructure, the approximation of a spherically symmetric structure for the radial profile derivation is reasonable given the weakness and distance of the secondary feature.

Figure~\ref{G205_fitting} shows that the radial profile of G205M3 flattens out beyond a radial distance of about 12000~au. 
The constant though noisy profile beyond 12000~au may be due to the ambient molecular cloud environment.
From fitting the SCUBA-2 column density profile, we find an outer radius of \similar~11000 au.
Combining all the estimations from different observing configurations, the best global fit values are  $n_c$ = 1.1\ee7 \cmq and $a$\,\similar 500~au. The global parameter values can be estimated from  the plot (see Appendix, Fig.~\ref{G205M3:dchi2}) -- the values corresponding to the common region of  $\Delta \chi^2$ contours.
The estimated prestellar core mass is 2.2 \msun, implying that roughly 30 percent of the emission was resolved out by the ACA observations \citep[mass 1.7 \msun, for a temperature of 10K and the same dust opacity, see][]{sahu2021}. 


  We note that for G205M3, the derived average number density within the flat region (500 au) is \similar 7\ee6 \cmq, about two times higher than  that of \similar $3.0$\ee6 \cmq toward L1544 found by \citet{keto2010MNRAS.402.1625K}.

\begin{table*}
   
    \centering
    \begin{tabular}{l l l l }
    \multicolumn{4}{c}{Core Parameters Determined by Different Methods}\\
    \hline
    \hline
    \multicolumn{4}{c}{G205M3}\\
    \hline
     &  $n_c$ (\cmq)  &  flat region size (a in au )  & radius (au)\\
    \hline
    Analytical model fit  & $1.35 \times 10^6$ (0.91) & 924 (410) & 10720 (1983) \\
     SCUBA cont fit*   & $3.4\times 10^7$ & 214.6  & - \\
     ACA cont fit   & $1.1 \times 10^7$ & 521 & - \\
     TM2+ACA cont fit   & $1.3 \times 10^7$ & 419  &- \\
     Global fit & 1.1 $\times 10^7$ &  521 & 
     -\\
     Plotted values & 1.1 $\times 10^7$ &  419, 521 & 
     -\\ 
     \hline
     \hline
     \multicolumn{4}{c}{ G209N}\\
     \hline
    Analytical model fit  & $6.08 \times 10^5$ (0.76) & 2891 (254) & 17178 (770) \\
     SCUBA cont fit   & $1.4\times 10^6$ & 2158.4  & - \\
    ACA cont fit   & $2.4 \times 10^6$ & 1749 & - \\
     TM2+ACA cont fit   & $3.6 \times 10^6$ & 1135  &- \\
     Global fit & 2.9 $\times 10^6$ &  1340 &     -\\
     Plotted values & 2.9 $\times 10^6$ &  1340 &     -\\
     \hline 
     \hline 
     \multicolumn{4}{c}{G212N1}\\

     \hline
    Analytical model fit  & $8.6 \times 10^5$ (7.18) & 1009 (548) & 7357 (214)  \\
     SCUBA cont fit   & $2.9\times 10^6$ & 726  & - \\
   ACA cont fit*   & $1.3 \times 10^6$ & 1544 & - \\
     TM2+ACA) cont fit   & $9.1 \times 10^6$ & 214  &- \\
    Global fit & 5.2 $\times 10^6$ &  316 &     -\\
    Plotted values & 5.2 $\times 10^6$ &  316, 521 &     -\\ 
     \hline

    \end{tabular}
    \caption{Fitting results for the three cores. The core radius obtained using the analytical fitting  are presented in row one with its error bar shown in the bracket. The best fit values obtained from grid search for different telescope configurations are presented in rows two to five;  note that the uncertainties are not noted in the table and visually presented in Appendix Section~C. The global fit represents the best fit parameter values considering multi-scale observations. For {\bf G205M3}, Fig.~\ref{G205_fitting} shows the profile plots with the 'global fit' parameter values for ACA and 12m-TM2 observations, while for the case of SCUBA-2, 'a' \similar 419 au is used. For {\bf G209N}, Fig.~\ref{G209N fitting} shows a similar plot with the  global parameter values obtained. In case of {\bf G212N1}, Fig.~\ref{G212 fitting} is a plot using global parameter values for the  TM2 configuration, while for ACA and SCUBA-2 case,  $n_c$ \similar $5.2\times 10^6$  cm$^{-3}$ and `a' \similar 521 au is used.
    }
    \label{table1}
\end{table*}


\subsection{G209N} 
The G209N core is comparatively larger and more elongated than G205M3, with no substructure present at the single dish scale (\similar 10000 au). In the higher resolution interferometric observations, unlike G205M3, G209N shows a dense central peak with a nearly circular shape.

\begin{figure*}
    \centering
    \includegraphics[width=16cm]{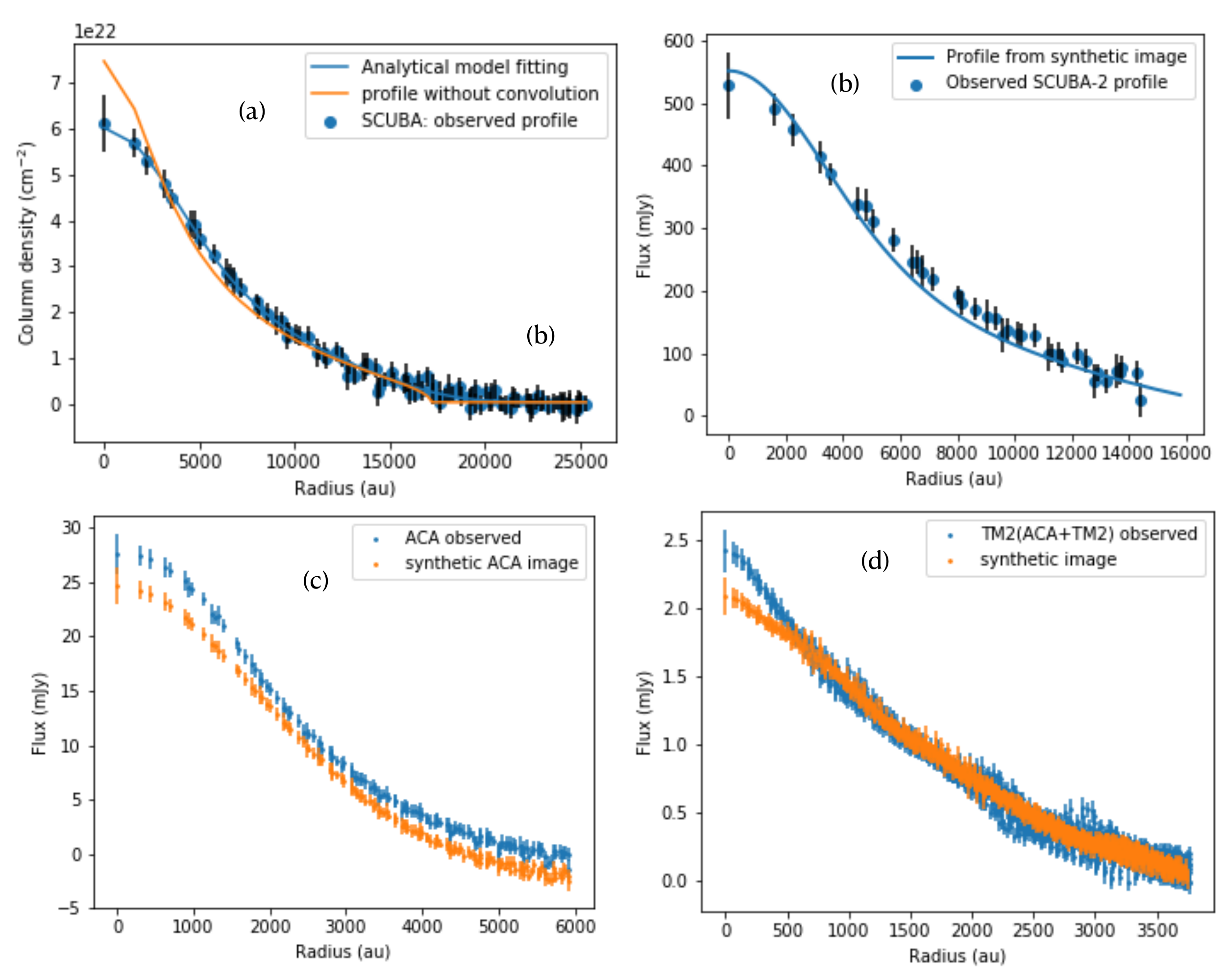}
    \caption{Similar to the fitting plots as Fig.~\ref{G205_fitting}; for the core { \bf G209N}.  The model parameters are noted in Table~\ref{table1}. A comparison between the modeled images with the observed maps are shown in Appendix  Fig.~\ref{G209_modelmap}.)}
    \label{G209N fitting}
\end{figure*}

Through an analysis similar to that for G205M3, the outer radius, $R$, of G209N is found to be 42\farcs5 or \similar 17000 au at a distance of 400~pc  (see Fig.~\ref{G209N fitting}).
The best fit parameters, using our multi-scale observations, are constrained to be {\bf $n_c$}\similar 2.9\ee6 \cmq and  $a$\similar 1300~au. These values are listed in Table~\ref{table1}  (also see Appendix, Fig.~\ref{G209N dchi2}).
  For G209N, the average number density  within the flat region \similar 2\ee6 \cmq  may be  slightly smaller or similar to that of L1544  \citep{keto2010MNRAS.402.1625K,Caselli2019ApJ...874...89C}. 
 The mass inferred for G209N is 10.5~\msun. Given its large outer radius, a significant fraction of emission was resolved out by the ACA, which detected a mass of only 2.7~\msun \citep{sahu2021}.

\subsection{G212N1}

Toward G212N1, a secondary peak can be seen \similar 25\arcsec\ to the SE in the SCUBA-2 map (Fig.~\ref{aca_scuba}).
Assuming that this nearby source, beyond 8500~au, is a separate entity, we focus on the densest substructure detected by ALMA (ACA and TM2 configurations) toward the field center (see Fig.~\ref{aca_tm2}).
The core radius, $R$, is found to be $\sim$ 7400~au from the column density profile fitting, and the profile plots related to each of the observing configurations are presented in Figure~\ref{G212 fitting}.

From the $\Delta \chi^2$ plots  (see Appendix,  Fig.~\ref{G212 dchi2}), there is no obvious common parameter region corresponding to the SCUBA-2, ACA, and TM2 observations, though the contours are  quite close and have partial overlap. 
Thus, the best-fit results for each different observational setup are noted in Table~\ref{table1}.  The peak density \similar 5.2\ee6 \cmq and the size of the flat region $a$ \similar 320 au corresponding to the global fit, do not simultaneously fit well the observed intensity profiles from different observations .  The small flat inner region is determined from the 12-m TM2 observation and the corresponding profile. Fixing the best-fit peak density \similar 5.2\ee6 \cmq, we find a somewhat larger flat region $a$ \similar 520 au for the ACA and SCUBA-2 observations. Within a reasonable uncertainty, a peak density \similar 5.2\ee6 \cmq and $a$ \similar 320 - 520 au closely describes the central zone across the multi-scale observations.

The core radius, $R$, of G212N1 is significantly smaller, \similar 7000 au, than for the other two Orion prestellar cores, and the dust continuum emission is more likely fully recovered by the ACA observations - the theoretically estimated core mass is \similar 0.95 \msun\ close to be the value of 1.0 \msun\ observed with the ACA \citep{sahu2021}.

\begin{figure*}
    \centering
    \includegraphics[width=16cm]{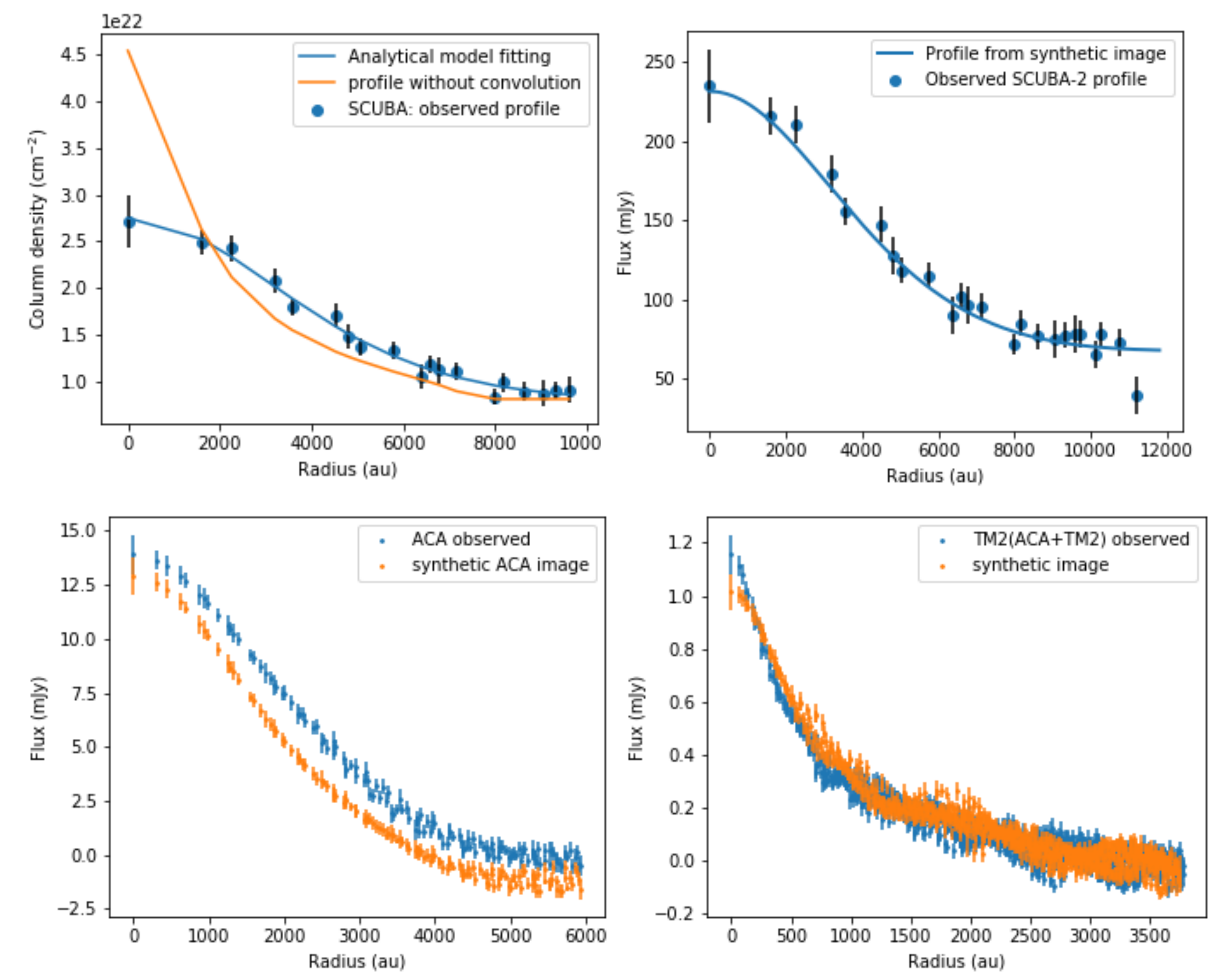}
    \caption{Similar to the fitting plots as Fig.~\ref{G205_fitting}; for the core G212N1.  The model parameters are noted in Table~\ref{table1}. A comparison between the modeled images with the observed maps are shown in Appendix  Fig.~\ref{G212_modelmap}.}
    \label{G212 fitting}
\end{figure*}

\section{Discussion} 

\subsection{Model uncertainties from the assumptions of temperature and dust opacity}

We have assumed an uniform temperature of 10 K for the cores both in the initial column density profile fitting for determining the outer radius and in the radiative transfer calculation for generating synthetic maps on the parameter grids. 
In the central region of the cores, however, the density is presumably higher and the temperature may be correspondingly lower as compared to the outer layers due to efficient dust continuum cooling.
Quantitatively, \citet{Crapsi2007A&A...470..221C} using high resolution interferometric observations of \ce{NH3} emission found a drop in the gas temperature of \similar 5 K, and therefore  well coupled dust temperature, in the centre of L1544.

\begin{figure}
    \centering
    \includegraphics[width=6.9cm]{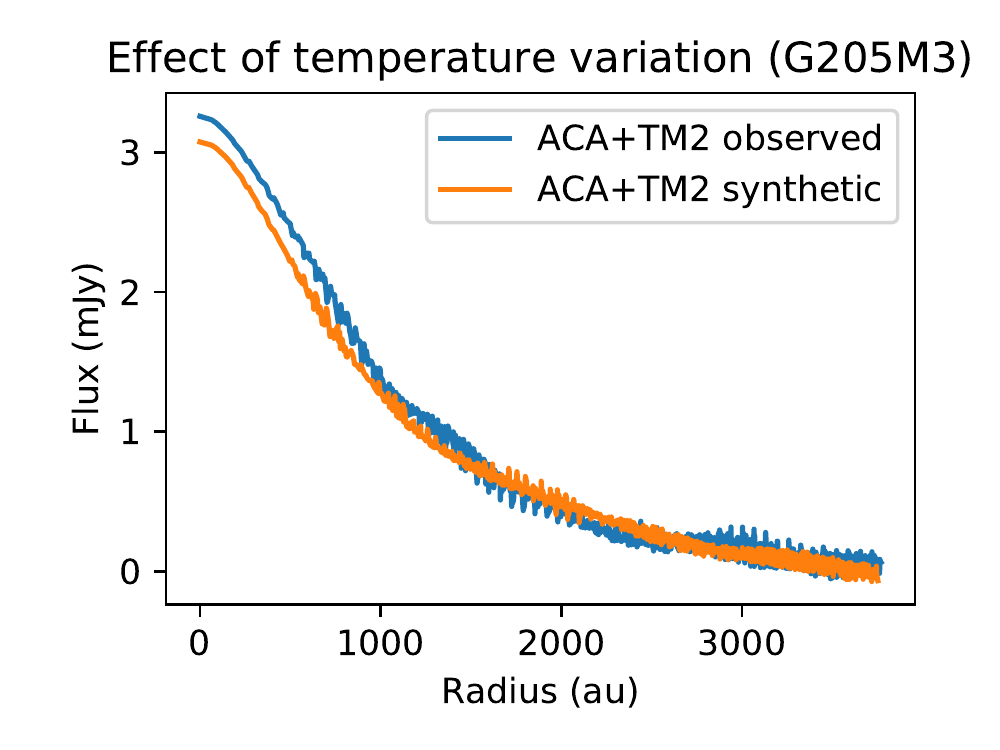}
    \caption{Effect of temperature variation for a example core G205M3. Here model image's $n_c$ and `a' are 1.9\ee7 \cmq and 316 au, respectively. These values can be compared with the values in Table~\ref{table1}, Row~4 }
    \label{profile_tvari}
\end{figure}

To check the effect of a varying temperature profile on the inferred model parameters, we consider the temperature profile given by \cite{Chacon2019}:  
\begin{equation}
    T_{out}= T_{out} - \frac{T_{out} - T_{in}}{1 + (\frac{r}{a})^{1.7}},
\end{equation}
estimated via a radiative transfer calculation for the prestellar core L1544. 
As the PSCs presented here are at least as centrally dense as L1544, we assume a similar temperature profile with $\mathrm{T_{in}}$= 6.9~K and $\mathrm{T_{out}}$ = 12~K \citep{Crapsi2007A&A...470..221C, Chacon2019}. 
Using this temperature profile and the same parameter space as studied earlier, we find that, for G205M3, the best fit peak density and flat radius (for 12-m TM2+ACA combined observations; compare with row 4, Table~\ref{table1}) are \approxt 1.9\ee7 \cmq and 320 au, respectively (Fig.~\ref{profile_tvari}). Thus, qualitatively, the results imply that the inferred peak densities may be slightly higher (\similar 50\%)  and the inner flat cores somewhat more compact (\similar 25\%). The estimated values of peak density and flat radius will be similarly affected for other cores too.

Another assumption in the imaging modelling process, in addition to an isothermal temperature, is the choice of dust opacity used.
For instance, \citet[][and references therein ]{keto2014MNRAS.440.2616K} were successful in reproducing the molecular line and continuum observation of L1544 by considering temperature variations \citep{Crapsi2007A&A...470..221C}, but only if an increased/scaled dust opacity was included.
Indeed, variations in dust opacity within  molecular clouds have been deduced in recent years  \citep[see, e.g.,][]{Sadavoy2016A&A...588A..30S, Juvela2015A&A...584A..93J}.
The reasons for such opacity behavior are not clearly understood and highly uncertain. 
Recently, however, \citet{Chacon2019} studied in detail the effect of opacity on the density profile of the prestellar core L1544. 
By considering a radial opacity variation, they found that the corresponding best-fit density profile is comparatively flatter than that derived by  \citet{Crapsi2007A&A...470..221C}.
Consequently, the peak density was also lower by a factor of 25\%. 
Applying a radial temperature variation, we have already shown that the peak density becomes higher by about 50\% and the source becomes more compact by around 25 \%.
Thus, guided by the results of \citet{Chacon2019}, we expect that our estimated parameters assuming isothermal and constant opacity conditions do not differ much from those if we consider the combined effect of temperature and opacity variations.

\subsection{Choice of the density profile}

Though their assumptions may differ, the theoretical models for star forming cores such as the Larson core, \citep{Larson1969MNRAS.145..271L}, the inside-out collapse of a singular isothermal sphere (SIS)\citep{Shu1977}, and the quasi-equilibrium (QE) BE sphere \citep{Bonnor1956} all suggest that the gas densities in the outer region of a spherical core can be asymptotically described by the $r^{-2}$ profile while the inner region exhibits a shallower radial dependence.
Even non-spherical models that consider the effect of ambipolar diffusion and large-scale turbulence produce similar $r^{-2}$ outer density profiles \citep{Ciolek2000ApJ...529..925C}.

Observational studies of starless/prestellar cores indeed seemed to find a central plateau in addition to power law density profile in the outer region \citep[e.g.,][]{Whitworth2001ApJ...547..317W}.
For more general cases, \citet{Shirley2000ApJS..131..249S}, for example, found that the density profiles of low-mass cores lie in the range of r$^{-1.5}$ to r$^{-2.6}$ from a survey of 21 cores, including starless and protostellar objects.
\citet{Hung2010} found two types of cores from their analysis of starless/prestellar cores --- the `steep type' which can be fit by a power law slope of $-2.5$, and the `shallow type' which can be described by a power-law slope $\sim -1.2$. 
Such more `exotic' shallow or steep power-law profiles may be resulted from the fact that either fitting the inner flat region together with the outer $r^{-2}$ profile by a single power-law or the object is more of prostellar instead of being starless in nature\citep[also see][]{Kwon2009ApJ...696..841K}.

  While protostellar cores are usually found to exhibit a power-law of r$^{-2}$ in their density profile \citep[e.g.,][]{Shirley2000ApJS..131..249S, Kwon2009ApJ...696..841K},
observational studies of starless/prestellar cores in contrast found that a central plateau in addition to a power law in the outer region indeed provides a better description for their density profile \citep[e.g.,][]{Whitworth2001ApJ...547..317W}.
For more general cases, \citet{Hung2010}, for example, found two types of cores from their analysis of starless/prestellar cores --- the `steep type' which can be fit by a power law slope of $-2.5$, and the `shallow type' which can be described by a power-law slope $\sim -1.2$. 
Their results suggested that the ‘steeper type’ starless cores are more evolved compared to the ‘shallow type’ of cores. \citet{Gomez2021MNRAS.502.4963G} indicated that the observed density profiles for starless cores with slopes steeper than -2.0 in fact could be mostly remedied by various observational uncertainties.  

Motivated by the above,  we adapted the proposition of \citet{Daap2009MNRAS.395.1092D}, an 
analytical density profile with an inner flat region and an outer r$^{-2}$ profile as described in Eq.~\ref{eq2}.
This model profile resembles that of the BE sphere but without the assumption of the core being in equilibrium.
In addition, the temperature is treated as a input parameter for the model, as is done for our purpose.
With the assumed r$^{-2}$ profile, we were able to better determine the outer boundary of the cores, which helps to reduce the degeneracies between the central density, the flat region size, and the core size for a given peak column density.



\subsection{Implications of the flat region model parameters} 

The model parameters of the flat region, including its central density and size, provide useful insights about the dynamical state of the core under consideration.
In their work of modeling of starless cores, \citet{Keto2008ApJ...683..238K} considered starless cores in two categories - those being thermally subcritical and thermally supercritical. The classification depends whether their central density is below or above a few $\times$ 10$^5$ \cmq, respectively.
The central densities of the cores presented here are much higher than a few $\times$ 10$^5$ \cmq, so the cores are arguably to be dynamically unstable and prone to dynamical collapse.
The peak central density and the small flat region of G205M3, for example, may furthermore correspond to roughly a young dynamical age of a few 10$^5$ years to 1 Myr in the model of \citet[][see their Fig. 1]{keto2010MNRAS.402.1625K}.

Indeed, \citet{Daap2009MNRAS.395.1092D} also indicated that,
as the core evolves, the core's central density increases and the size of the flat region reduces.
The inner flat region appears during the core evolution when the sound crossing time is less than the free-fall time and any density perturbations are rapidly smoothed up by the pressure waves within.
Following this logic, the flat region size $a$ value can be estimated as the product of sound speed, which depends on the gas temperature (which is assumed to be a constant here), and the free-fall time, which inversely depends on the square root of the mean density of the enclosed mass and therefore drops as the central density increases.
More explicitly, the flat region size can be expressed as $a= k \frac{C_s}{\sqrt(G \rho_c)}$, where $k$ is the constant of proportionality and the latter part is the Jeans length. 
The analytical estimation from \citet{Daap2009MNRAS.395.1092D} shows that for a BE sphere, the core would be collapsing (i.e., supercritical) if $k \sim 0.6$ or higher. 
In the case of G205M3, with a flat region size of \approxt 500 au and a peak density of \approxt 1.1\ee7 \cmq, the $k$ value is \similar 0.7. Similarly, for G209N, $k$ value is 0.9. 
If the temperature of the inner region is lower than 10 K, the $k$ value would be even higher than those estimated values. 
Therefore the cores, G205M3 and G209N are likely to be collapsing to form protostar/protostars.
For, G212N1, the value of $k$\similar 0.5 is at the borderline for a collapsing core and this core may thus be on the verge of collapse.
Nevertheless, the set of prestellar cores represent the most compact (smaller flat region size) PSCs known to date, and future spectral line observations toward the sample may provide direct evidence of collapsing signatures.

\section{Conclusion}
 
 From our past ALMA observations we have found five highly dense and centrally compact PSCs, namely, G205.46-14.46M3, G208.68-19.20N2, G209.29-19.65S1, G209.94-19.52N,  and G212.10-19 15N1 in Orion.
Among the five PSCs, we found  three - G205.46-14.46M3, G209.94-19.52N,  and G212.10-19 15N1 (in short G205M3, G209N1, G212N1 respectively)  are suitable to estimate the density profile with the assumption of underlying  spherical shape. We further studied in detail the density profile of the three cores, considering the multiscale observations with beam sizes ranging from 5600 au (single dish SCUBA-2) to 480 au (ALMA). To reproduce these cores' density profile, we  consider a  physical model which includes a flat central region at the center and \rsq profile outside. The cores were assumed to be isothermal at 10~K.
For two cores, G205M3 and G209N1, we found that the density profile and dust continuum observation from the multiscale observations to be closely consistent with the assumed physical model. A slight deviation from the theoretical profile  may be present for the third core G212N1, which has a prominent substructure at a large scales ($\sim$ 10000 au). The peak density, flat region size and outer radius of the cores are found to be \approxt  ( 1\ee7 \cmq, 500 au, 11000 au), 
( 3\ee6 \cmq, 1300 au, 17000 au) and ( 5\ee6 \cmq, 300 au, 7000 au), respectively, for G205M3, G209N and G212N1. 
Though we do not consider variations in core temperature and dust opacity, we found that these factors do not significantly affect the estimated peak density and flat region size. We used the estimated flat region size to gauze the dynamical state of the PSCs. 
The sizes of the flat regions imply that the cores are unstable and gravitationally collapsing or on the verge of collapse.
Based on the estimated peak density and flat radius sizes, the cores in our current study belongs to  a rare sample of highly evolved prestellar cores known to date.

\acknowledgments
We thank Paola Caselli for her careful review and helpful suggestions. This paper makes use of the following ALMA data: ADS/JAO.ALMA\#2018.1.00302.S. ALMA is a partnership of ESO (representing its member states), NSF (USA) and NINS (Japan), together with NRC (Canada), MOST and ASIAA (Taiwan), and KASI (Republic of Korea), in cooperation with the Republic of Chile. The Joint ALMA Observatory is operated by ESO, AUI/NRAO and NAOJ.
DS acknowledges the support from SERB, Govt. of India and PRL (A Unit of DoS) for continuing research and capacity building through the Ramanujan (Faculty) Fellow position. DS also acknowledges the support from Academia Sinica and TIARA (ASIAA) computing facilities.  
SYL acknowledge support from the National Science and Technology Council (NSTC) with grants 110-2112-M-001 -056 - and 111-2112-M-001 -042 -. DJ is supported by NRC Canada and by an NSERC Discovery Grant. 
Tie Liu is supported by the National Key R\&D Program of China (No. 2022YFA1603100), National Natural Science Foundation of   China (NSFC) through grants No.12073061 and No.12122307, the international partnership program of Chinese Academy of Sciences through grant No.114231KYSB20200009, Shanghai Pujiang Program 20PJ1415500 and the science research grants from the China Manned Space Project with no. CMS-CSST-2021-B06.  N.H. acknowledges NSTC 110-2112-M-001-048 and NSTC 111-2112-M-001-060 grants. M.J. acknowledges support from the Academy of Finland grant No. 348342.  C.W.L. is supported by the Basic Science Research Program through the National Research Foundation of Korea (NRF) funded by the Ministry of Education, Science and Technology (NRF- 2019R1A2C1010851), and by the Korea Astronomy and Space Science Institute grant funded by the Korea government (MSIT; Project No. 2022-1-840-05). LB gratefully acknowledges support by the ANID BASAL projects ACE210002 and FB210003. W.K. was supported by the National Research Foundation of Korea (NRF) grant funded by the Korea government (MSIT) (NRF-2021R1F1A1061794).
 This research was carried out in part at the Jet Propulsion Laboratory which is operated by the California Institute of Technology under a contract with the National Aeronautics and Space Administration (80NM0018D0004).

\appendix
\restartappendixnumbering
\section{The five prestellar cores }

\begin{figure*}
\centering
    \includegraphics*[width=18cm]{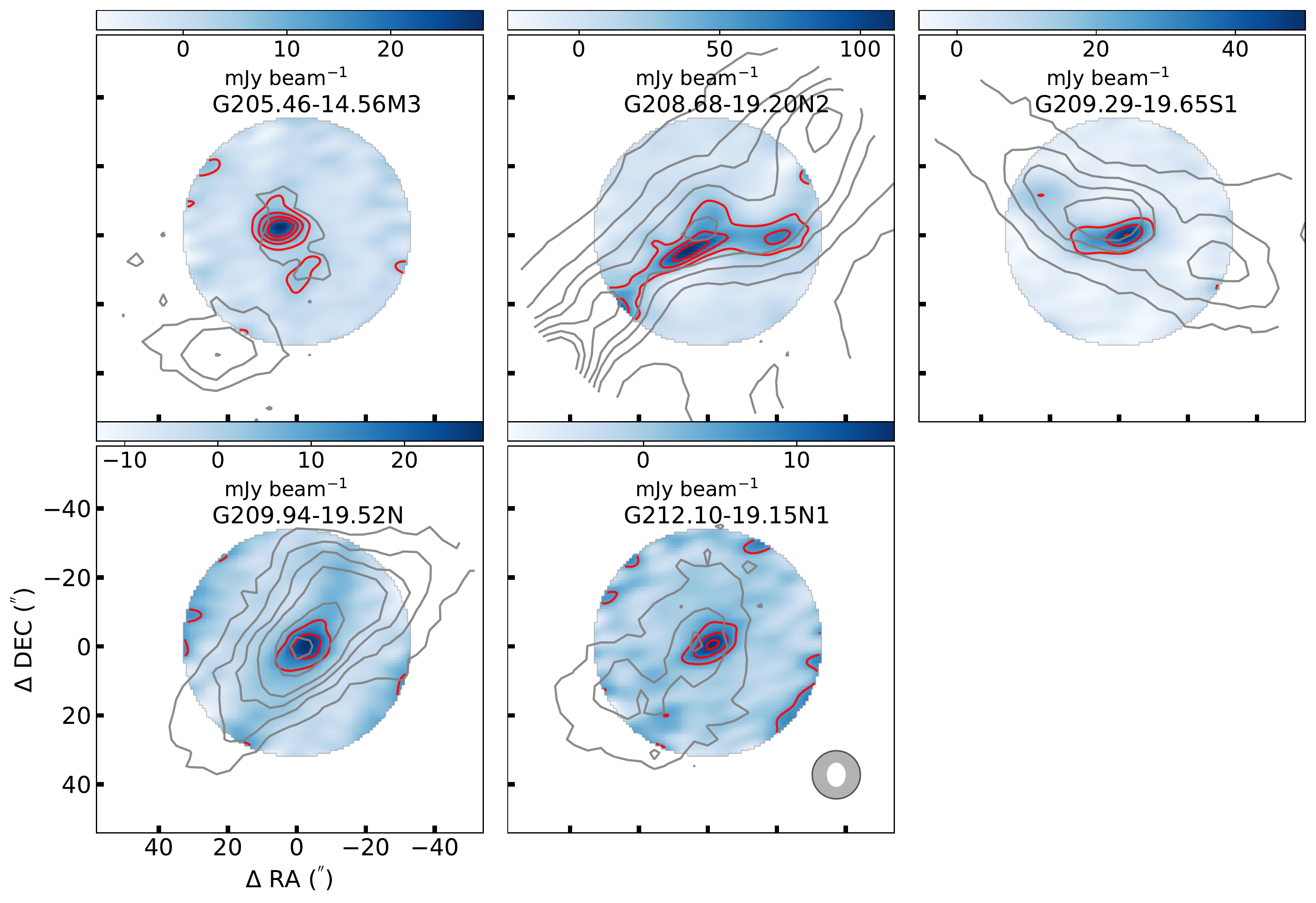}
    \caption{The 1.3 mm dust continuum emission of the five prestellar cores( in blue colors); red contours correspond to 5, 10, 15$\sigma$, where $\sigma$(rms) are 1.0, 6.0, 3.3, 2.0, 1.0 mJy~beam$^{-1}$ respectively for the cores G205M3, G208N2, G209N1, G209N, G212N1. The grey contours correspond to 0.85mm SCUBA-2 emission with similar contour levels where  $\sigma$(rms) =44.5, 64.2, 31.3, 13.9, 16.4 mJy~beam$^{-1}$ respectively. The beams of SCUBA-2 and ACA are shown in the lower-right panel. } 
    \label{aca_scuba_all}
\end{figure*}

In the article, we  discussed only the three cores among the five sample cores that were detected by \citet{sahu2021};  Figure~\ref{aca_scuba_all} shows them all. Two of the five sample cores, G208N2 and G209S1,  are very elongated and have asymmetric structures  and often treated as filaments \citep{Ohashi2018ApJ...856..147O}. So, the radial profile estimation on the assumption of spherically symmetric core is likely unsuitable in these cases, and excluded from the analysis.

\section{Comparison of emission maps}
\restartappendixnumbering
In this section, we show the observed dust continuum vs. synthetic model images  of the three cores that are presented in the text.
 \begin{figure*}
   \centering
    \includegraphics[width=18cm]{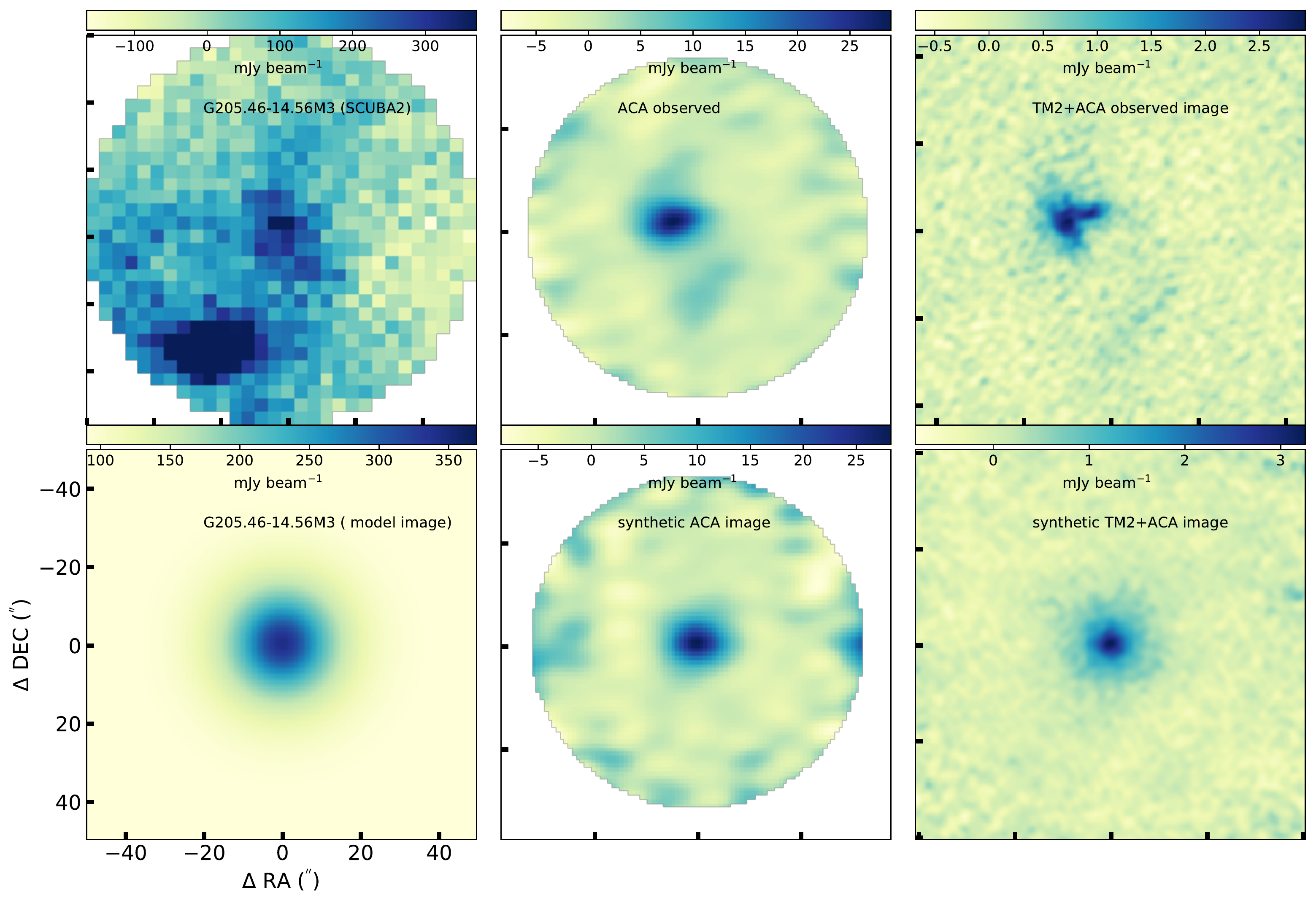}
    \caption{Upper panels show the observed dust continuum from SCUBA-2, ACA and TM2, respectively towards G205M3 core. The corresponding  modeled synthetic images are shown in the lower panels. The model images are based on the best fit results that are presented in  Fig.~\ref{G205_fitting} and Table~\ref{table1}.
    }
    \label{G205_modelmap}
\end{figure*}

\begin{figure*}
   \centering
    \includegraphics[width=18cm]{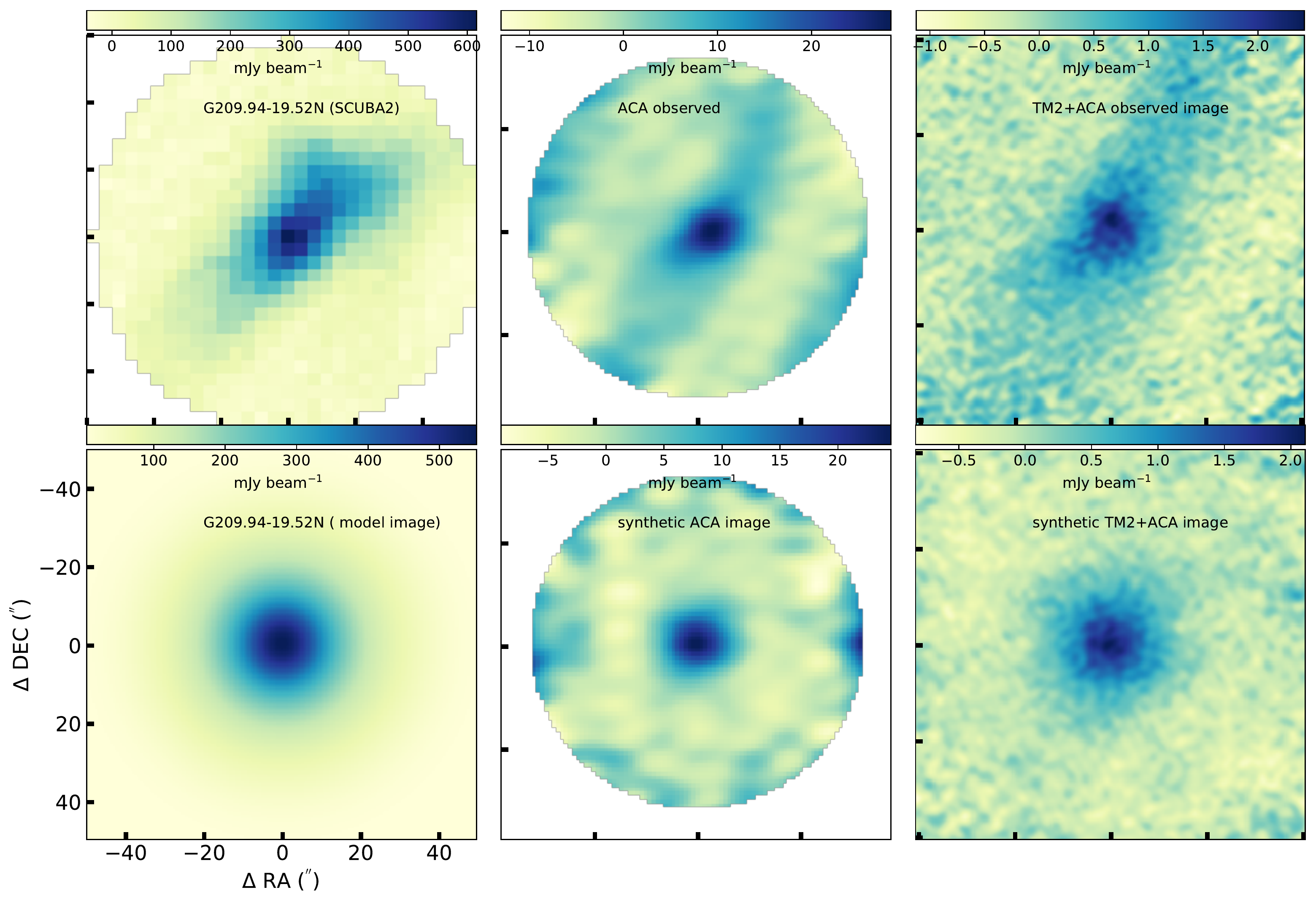}
    \caption{Upper panels show the observed dust continuum from SCUBA-2, ACA and TM2, respectively towards G209N core. The corresponding  modeled synthetic images are shown in the lower panels. The model images are based on the best fit results that are presented in  Fig.~\ref{G209N fitting} and Table~\ref{table1}}
    \label{G209_modelmap}
\end{figure*}

\begin{figure*}
    \centering
    \includegraphics[width=18cm]{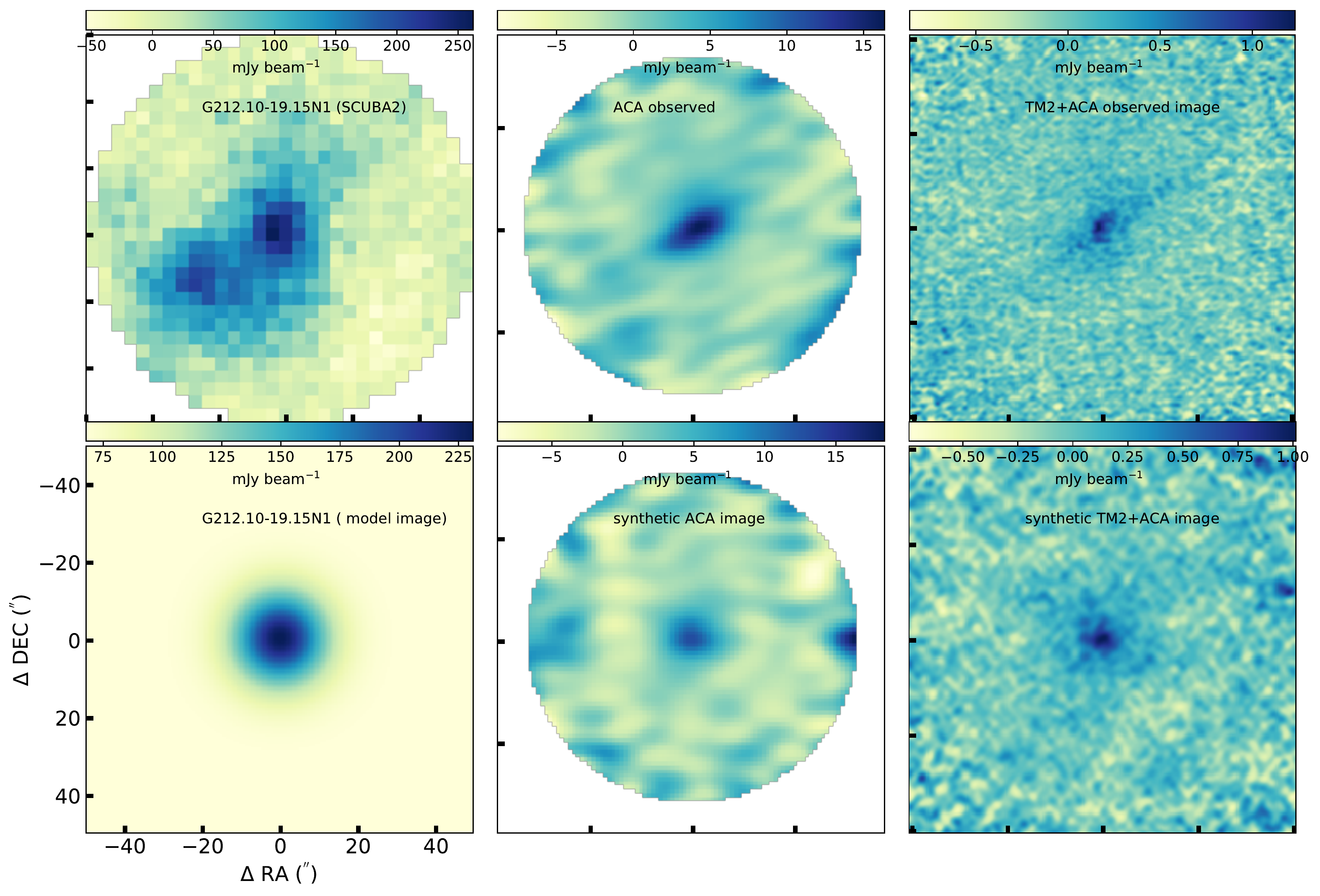}
    \caption{Upper panels show the observed dust continuum from SCUBA-2, ACA and TM2, respectively towards G212N1 core. The corresponding  modeled synthetic images are shown in the lower panels. The model images are based on the best fit results that are presented in  Fig.~\ref{G212 fitting} and Table~\ref{table1}.}
    \label{G212_modelmap}
\end{figure*}

\section{ $\Delta \chi^2$ and uncertainity estimation }

\restartappendixnumbering
\begin{figure}
    \centering
    \includegraphics[width =9cm]{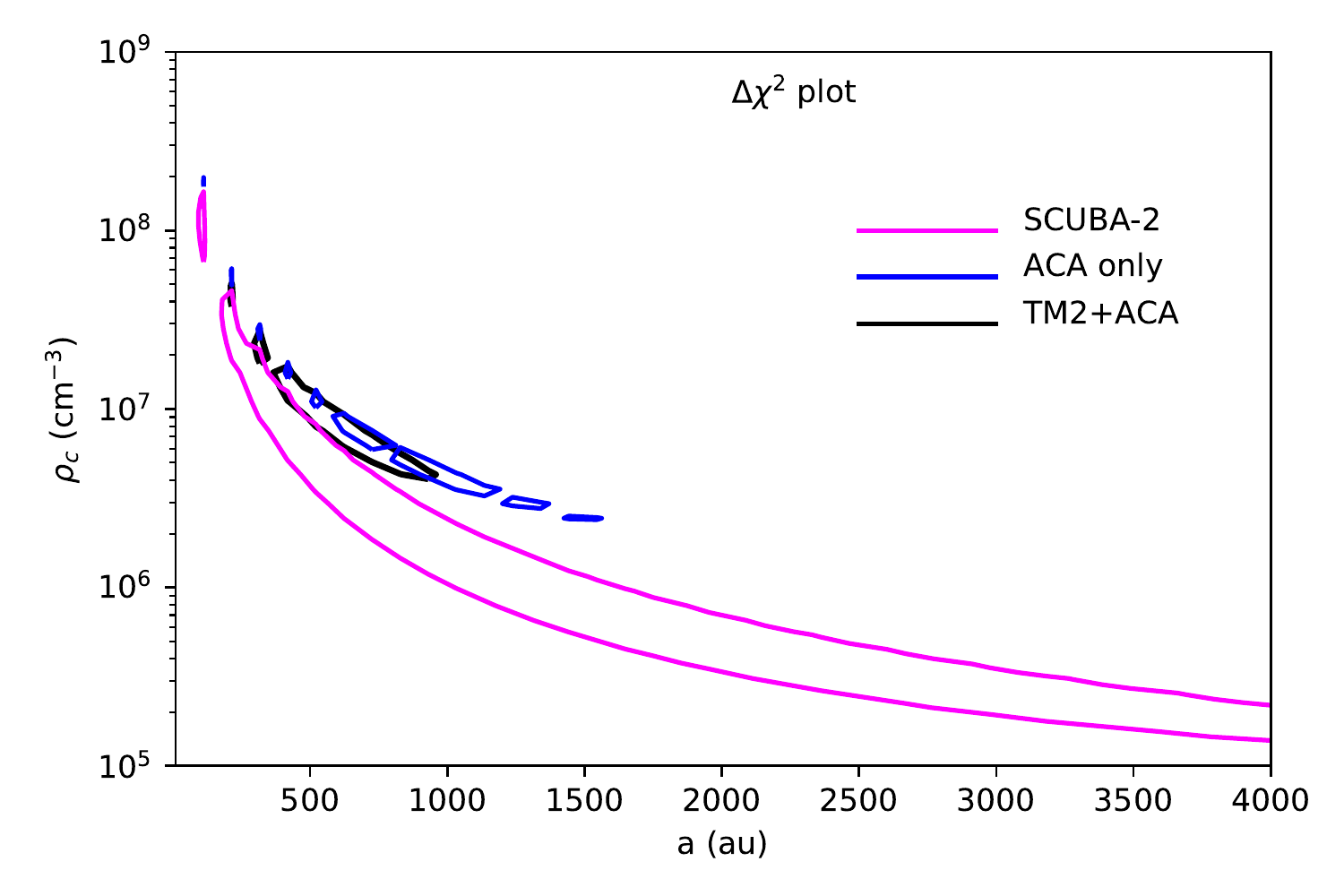}
    \caption{ $\Delta~ \chi^2$ plot as obtained from the parameter space used to fit the multiscale observations from SCUBA-2 to ALMA for the {\bf G205M3 core} }
    \label{G205M3:dchi2}
\end{figure}

\begin{figure}
    \centering
    \includegraphics[width=9cm]{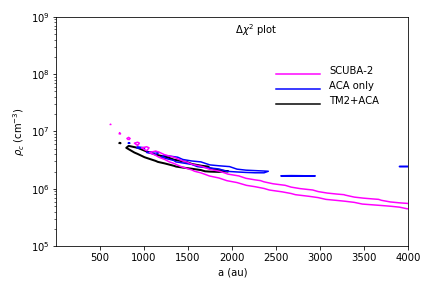}
    \caption{$\Delta~ \chi^2$ plot for the {\bf G209N} core, similar to Fig.~\ref{G205M3:dchi2} }
    \label{G209N dchi2}
\end{figure}

\begin{figure}
    \centering
    \includegraphics[width=9cm]{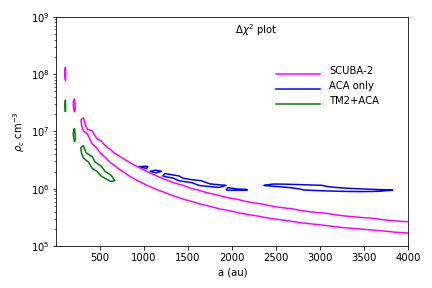}
    \caption{$\Delta~ \chi^2$ plot for the {\bf G212N1} core, similar to Fig.~\ref{G205M3:dchi2} }
    \label{G212 dchi2}
\end{figure}

\bibliography{sample63}{}
\bibliographystyle{aasjournal}



\end{document}